# Development and first operation of a Cavity Ring Down Spectroscopy diagnostic in the negative ion source SPIDER

M. Barbisan,[1,a] R. Pasqualotto,[1] R. Agnello,[2] M. Pilieci,[3] G. Serianni,[1] C. Taliercio,[1] V. Cervaro,[1,b] F. Rossetto,[1] and A. Tiso[1]

[1] *Consorzio RFX, Corso Stati Uniti 4 – 35127 Padova, Italy*
[2] *Ecole Polytechnique Fédérale de Lausanne (EPFL), Swiss Plasma Center (SPC), CH-1015 Lausanne, Switzerland*
[3] *Università degli Studi di Padova, Via 8 Febbraio, 2 - 35122 Padova*

The Neutral Beam Injectors of the ITER experiment will rely on negative ion sources to produce 16.7 MW beams of H/D particles accelerated at 1 MeV. The prototype of these sources was built and is currently operated in the SPIDER experiment (Source for the Production of Ions o Deuterium Extracted from an RF plasma), part of the Neutral Beam Test Facility of Consorzio RFX, Padua. In SPIDER, the $H^-/D^-$ ion source is coupled to a three grids 100 kV acceleration system. One of the main targets of the experimentation in SPIDER is to uniformly maximize the extracted current density; to achieve this it is important to study the density of negative ions available in proximity of the ion acceleration system. In SPIDER, line-integrated measurements of negative ion density are performed by a Cavity Ring Down Spectroscopy (CRDS) diagnostic. Its principle of operation is based on the absorption of the photons of a laser beam pulse by $H^-/D^-$ photo-detachment; the absorption detection is enhanced by trapping the laser pulse in an optical cavity, containing the absorbing medium (i.e. negative ions). The paper presents and discusses the CRDS diagnostic setup in SPIDER, including the first measurements of negative ion density, correlated to the main source parameters.

## I. INTRODUCTION

The ITER experiment foresees various additional heating systems, among which two (with provision for three) Neutral Beam Injectors (NBIs) will have to deliver beams of H/D particles at 870 keV (H) / 1 MeV (D), for a power of about 17 MW per NBI.[1,2] The most efficient way to produce a beam of neutral particles at this energy is to neutralize a beam of $H^-/D^-$ ions, generated inside an radiofrequency (RF) inductively coupled plasma source and extracted and accelerated by a system of electrically biased grids. ITER NBIs are required to extract negative ions with challenging current density values of about 330 A/m$^2$ (H) /285 A/m$^2$ (D) for one hour; to attain this, negative ions will be mostly produced by surface reactions, boosted by evaporation in the source of Cs, which lowers the surfaces work function.[1,2] To reach the requirements for the ITER NBI negative ion source against the several technical and physical challenges, a prototype is under study in the SPIDER (Source for the Production of Ions of Deuterium Extracted from an RF plasma) experiment, part of the Neutral Beam Test Facility of Consorzio RFX in Padova.[2-4] The SPIDER negative ion source is composed by eight RF plasma sources, also called "RF drivers", placed in a 2(h)x4(v) matrix;[1,2] the plasma expands from the drivers to a common expansion chamber, towards a three grids system (from the source: Plasma grid – PG, Extraction Grid – EG,

[a] Author to whom correspondence should be addressed: marco.barbisan@igi.cnr.it.
[b] Deceased, February 2021.

Grounded Grid – GG) accelerating negative ions up to 100 keV. To reduce the amount of co-extracted electrons, the source body and the Bias Plate (BP), which surrounds the PG apertures, can be independently electrically biased with respect to the PG. Besides this, a current of the order of $10^3$ A vertically flows through the PG to create an horizontal magnetic filter field, which reduces the electron temperature and density in proximity of the PG and consequently the fraction of co-extracted electrons.

The plasma source operation must be optimized in order to ensure the maximum and most uniform $H^-/D^-$ density in front of the PG apertures. Measuring the negative ion density also allows to verify that the $H^-/D^-$ beamlets are extracted and accelerated from each grids aperture with no vignetting.[5] To perform $H^-/D^-$ density measurements in the proximity of the PG in SPIDER, the Cavity Ring Down (CRDS) diagnostic was designed, and has been recently installed and operated. CRDS allows Line of Sight (LoS) integrated measurements of $H^-/D^-$ density, in a range from few $10^{15}$ m$^{-3}$ to about $10^{17}$ m$^{-3}$. Although CRDS has been extensively studied on many plasma devices,[6-11] its application on a full scale ion source for fusion, such as SPIDER, presented several technical issues to be solved;[12] the CRDS optic setup had to be made compatible and alignable with respect to the SPIDER structure, and stable enough to remain aligned under harsh mechanical conditions (vibrations due to the pumping system, thermal deformations, etc.). This paper will briefly recall the CRDS principle of operation (sec. II) and then the specific design and implementation in SPIDER (sec. III). The CRDS signals and a first characterization of negative ion production in SPIDER with respect to some basic source parameters will be shown.

## II. PRINCIPLE OF OPERATION

The core of the CRDS diagnostic[8] are two plano-concave high reflectivity (~99.99 %) mirrors, forming an optical cavity in axis with the desired LoS, having the region with H⁻/D⁻ in between. A laser beam is sent to the back side of the "input" cavity mirror, through which about $10^{-4}$ of the beam intensity enters the cavity and is reflected back and forth by the two cavity mirrors. The time evolution of the laser light stored inside the optical cavity depends both on the mirrors' reflectivity and on the presence of an absorbing medium. Laser photons can indeed turn H⁻ ions into neutrals: H⁻ + ν → H + e. The energy of the laser photons must exceed 0.75 eV, corresponding to the binding energy of the excess electron in the negative ion; the dependency of the photo-detachment cross section on photon energy is reported elsewhere.[13] On the opposite side with respect to the entrance of the laser beam, at each reflection on the "output" cavity mirror a small fraction ($10^{-4}$) of the beam exits. Since a pulsed laser is used, on the whole the output from the cavity consists of a train of light pulses, whose envelope shows an exponential decay. The cavity ring down time (or simply decay time) depends on the cavity length $L$, the path length d inside the H⁻/D⁻ region, the mirrors reflectivity $R$, the speed of light $c$, the negative ion density $n_{H^-}$ and the photo-detachment cross section $\sigma$. $n_{H^-}$ is obtained from the measured cavity ring down time $\tau$, compared to the cavity ring down time $\tau_0$ measured in absence of H⁻/D⁻ (i.e. with the source plasma switched off):

$$n_{H-} = \frac{L}{\sigma c d}\left(\frac{1}{\tau} - \frac{1}{\tau_0}\right) \qquad (1)$$

The minimum measurable value of negative ion density depends on the minimum measurable variation of τ.

## III. DIAGNOSTIC SETUP IN SPIDER

In SPIDER, the laser pulses for the CRDS diagnostic are provided by a *Quantel Q Smart 450* Q-switched Nd:YAG laser, with 1064 nm wavelength, emitting 150 mJ energy and about 6 ns duration pulses at 10 Hz. The photo-detachment cross section in eq. 1 is σ=3.5·10⁻²¹ m² at 1064 nm.[13] As shown in fig. 1, the laser is placed in a concrete shielded room outside the SPIDER bioshield, for protection from gamma and neutron radiation. On the same optical board on which the laser head is installed, an optical isolator protects the laser head from back reflections coming from the input high reflectivity (HR) mirror of the cavity; these reflections might cause severe damages to the laser head active rod. To simplify the alignment of CRDS optics, the infrared laser pulses are co-aligned with the continuous visible light of a diode laser, emitting 0.9 mW at 532 nm. Apart from the beam combiner, all the other mirrors are reflective for both 1064 nm and 532 nm and optimized for 45° incident angles. The mirrors and the laser diode support provide for the mechanical degrees of freedom necessary to co-align the two laser beams and to send them through an aperture in the bioshield, to reach the SPIDER vacuum vessel.

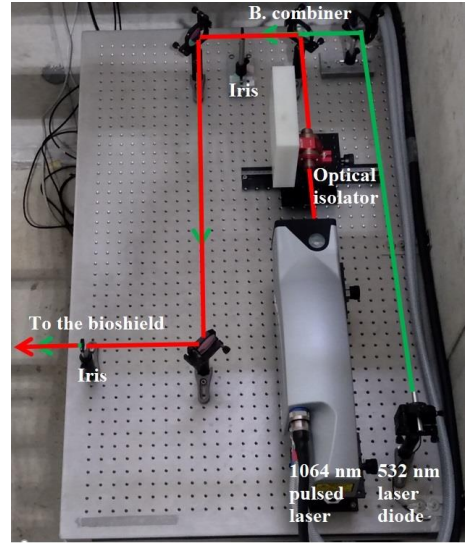

FIG. 1. Layout of the optical components in the CRDS room outside SPIDER bioshield.

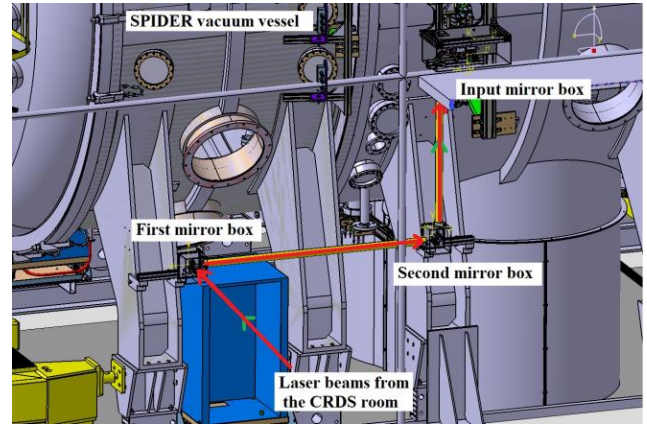

FIG. 2. 3D view of the optics installed in the bioshield to aim the laser beams towards the entrance of the optical cavity.

The laser beams, aiming towards the vessel, are deviated by two mirrors enclosed in protection metal boxes (first and second mirror boxes in fig. 2) in order to reach the optical cavity entrance. This is contained in the input mirror box, which hosts the optics required to properly match beams position and direction with respect to the optical cavity. A view from the bottom of the input mirror box optics is given in fig. 3. The mirror on the left brings the beams back to the horizontal plane. In future the mirror is envisaged to be replaced by a beam splitter, so that the beams will be available for up to four optical cavities/LoSs in total, at different vertical positions. A 3 m focal length plano-convex lens corrects the beams divergence for a better matching with the cavity Gaussian resonant mode $TEM_{00}$ at the cavity input.[12]

The mirror downstream the lens steers the laser beams to the correct position and aiming towards the input high reflectivity (HR) cavity mirror. This is a dielectric plano-convex mirror, with 25.0 mm diameter and 6 m radius of curvature for proper cavity stability; its reflective coating has a nominal reflection coefficient greater than 99.994% at the infrared laser wavelength and equal to 10 % at the alignment laser wavelength (532 nm). The mirror is an air vacuum interface. The cavity is L=4637 mm long and passes through 10 mm diameter holes drilled in the source. To obtain this the HR mirror is placed on a Viton O-ring, which ensures vacuum holding. The mirror can be tilted by means of three screws with 0.25 mm thread pitch, acting on the mirror on three points, 120° apart. The compression of the Viton ring allows an asymmetric deformation of it. This system is mounted on a CF40 flange, which can be translated ±10 mm both in horizontal and vertical direction thanks to a bellows and to two dovetail rails, stacked one on the other in perpendicular directions. The mirrors vacuum system is completed by a CF40 gate valve and a vent pipe; they will allow the replacement of a mirror with a worn coating without compromising the source vacuum.

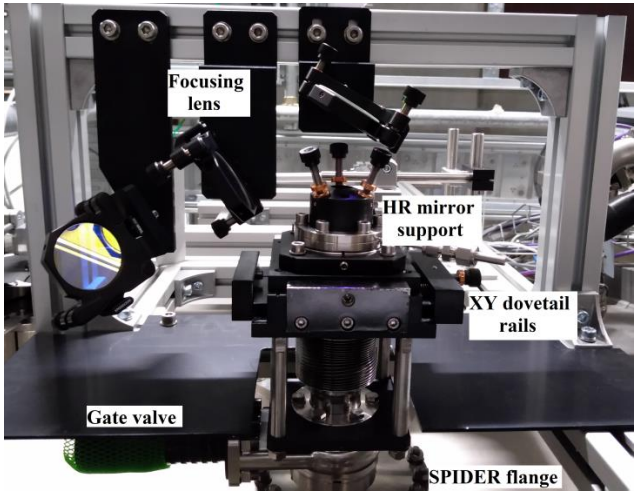

FIG. 3. View from the bottom of the optical components installed inside the input mirror box.

The optical cavity path position is shown in fig.4. The figure shows the Plasma Grid with its 16 groups of apertures. The PG has 1280 apertures for beam extraction, even though only 80 of them, marked by yellow spots, are currently available;[3] this is to minimize the gas flow from the source, due to pumping technical issues. As shown, the CRDS LoS/cavity travels on the top part of the four bottom aperture groups. The LoS is 5 mm upstream the PG surface, as close as possible, in order to study the future effects on negative ion production of Cs deposition on the PG. The CRDS laser path crosses the entire source-PG width, however the plasma and consequently the negative ions are present only in part of this path. In the source, indeed, a bias plate (depicted in light blue in fig. 4) is located at 10 mm in front of the PG surface;[2] its purpose is to create an electric field in proximity of the PG apertures in order to reduce the amount of co-extracted electrons. The CRDS line of sight is in between the PG and the bias plate, which is open to the plasma volume only in front of the groups of PG apertures (fig. 4). Determining the exact length of the LoS which is crossed by the plasma is not straightforward. At the moment, as a first order approximation, this length d is assumed to be 612 mm, i.e. the average between the width of the plasma chamber (720 mm) and the length over which the laser beam is under direct exposure of the plasma (504 mm).

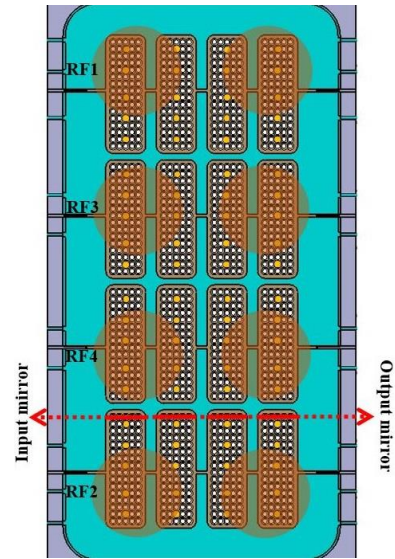

FIG. 4. Position of the CRDS LoS with respect to plasma grid apertures and bias plate (depicted in blue). The active apertures are marked by yellow spots. The large orange circles are projections of the radiofrequency drivers exits.

At the opposite side of the optical cavity, the output HR mirror is supported by a vacuum structure similar to the one of the input mirror (fig. 3), but with the additional support for a commercial collimator (0.25 numerical aperture) which conveys the light on a 1 mm core silica-silica fiber. The fiber leads the light in a separate building, in the optical room. There the light exiting from the fiber is re-focused by an aspheric condenser lens towards the detector active surface. An interference filter with 10 nm FWHM passband at 1064 nm is interposed between lens and detector, in order to block most of the light emitted by the plasma and collected by the collimator. The detector is a Hamamatsu S11519-30 avalanche photodiode, with about 70 A/W response at 1064 nm and about $10^2$ internal gain, coupled to a transimpedance amplifier with $4 \cdot 10^3$ V/A gain. For each laser pulse the digitizer, a CAEN DT5720A, acquires the detector output at 250 MS/s for a time interval of 1450 μs.

The main challenge in commissioning the diagnostic was aligning the optical cavity. This operation was initially performed with the visible laser only. The fine alignment was made possible by the 10 % reflectivity of the HR mirrors at 532 nm: the tilting of the mirrors was regulated to make the visible reflections coincide with the incident laser on each side. The final alignment was then performed using the infrared pulsed laser.

## IV. FIRST RESULTS

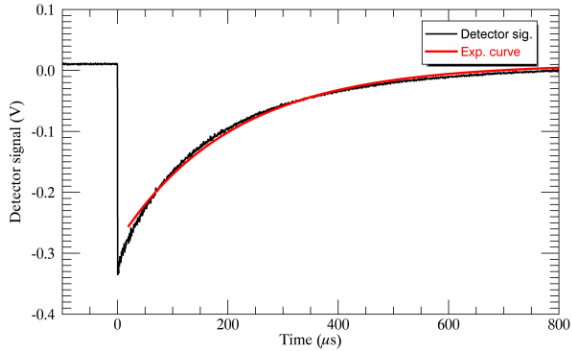

FIG. 5. Example of CRDS signal acquired by the diagnostic, as function of time. The fitted exponential curve is shown in red. In this case, $\tau=173$ μs.

The signals acquired by the detector did not show significant beatings between different cavity modes;[12] however the exponential fits, of type $y = b - A \cdot \exp(-t/\tau)$, where b is the background level, A is the signal amplitude, t is time and $\tau$ is the decay time, showed that the first 20 μs of each ring down clearly depart from an exponential decay. This likely suggests a multi-exponential decay given by several cavity modes, which rapidly disappear to leave $TEM_{00}$, the mode with the smallest transverse dimensions, alone. After the analysis of the first acquired CRDS data, it was empirically chosen to exclude the first 20 μs of each ring down signal from the exponential decay fit. A typical CRDS signal is provided in fig. 5, together with the exponential decay fit curve, indicated in red.

SPIDER plasma pulses are programmed so that the CRDS diagnostic starts firing the NIR laser at 10 Hz rate from t=-6 s from the plasma ignition, up to several seconds after the plasma switches off. This is useful not only to calculate a value of $\tau_0$ for each pulse, but also to correct a drift in the ring down time which occurs during a shot. One example is in fig. 6, showing the time evolution of ring down time during a 10 s plasma discharge in deuterium, in which the total RF power, also plotted in the figure with a red curve, was linearly ramped from 4x30 kW to 4x80 kW. Source pressure was 0.22 Pa, while the magnetic filter field current was 750 A. As seen in figure 6, $\tau$ exhibits a drift, quite linear in this case. The cause of this drift is still under investigation; thermal drifts are to be excluded, since the vessel (and CRDS optics mounted on it) are separated from the source body. The study of $\tau_0$ values from several SPIDER pulses led to the hypothesis that this drift may be due to a slow thermalization of the infrared laser internal optics. For the moment, for plasma pulses of duration up to few hundreds seconds, a linear fit is applied over those values of $\tau$ which correspond to "plasma off" times. These time intervals can be easily identified by means of the plasma light diagnostic.[14] From the resulting linear fit (magenta line in fig. 6) it is possible to calculate $\tau_0$ for any moment during the "plasma on" phase.

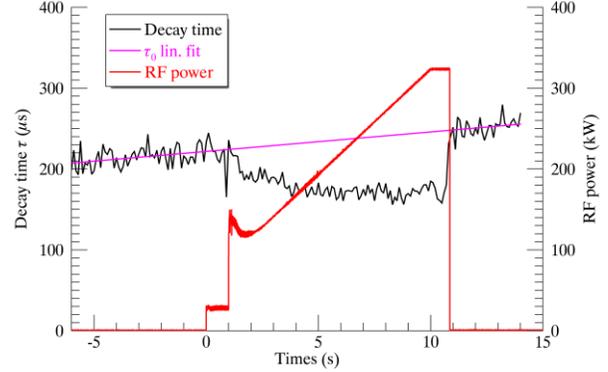

FIG. 6. Black curve: decay time before, during and after a plasma discharge in SPIDER. The magenta line indicates the linear fit for the estimation of $\tau_0$. Red curve: total radiofrequency power as function of time. Source pressure was 0.22 Pa; $I_{PG}$ filter current was 750 A.

Fig. 6 also shows that the $\tau$ measurements are prone to detectable fluctuations. While the mechanical vibrations of SPIDER vessel have been excluded and the laser pointing stability resulted to be below 50 μrad (i.e. oscillations of about 0.5 mm at the cavity entrance), it was still not possible to identify the cause of this phenomenon. The diagnostic is in any case able to detect values of negative ion density down to about $5 \cdot 10^{15}$ m$^{-3}$ or even less, averaging over multiple measurements.

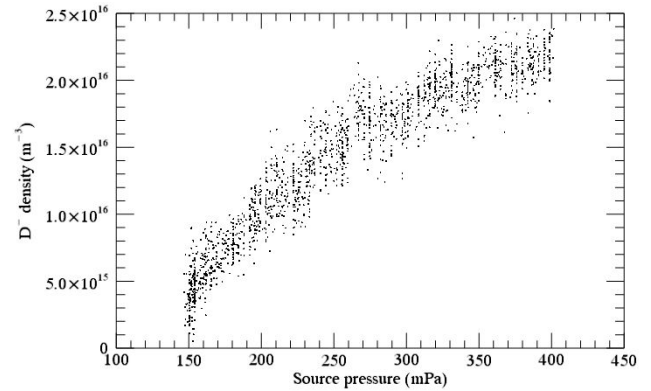

FIG. 7. D$^-$ density measured as function of source pressure. The plasma was sustained by 4x60 kW. $I_{PG}$ filter current was 750 A.

During the SPIDER experimental campaign in deuterium in October 2020 the negative ion production in Cs-free plasmas was studied. First, it was possible to characterize the dependency between negative ion density and source pressure, which as shown in Fig. 7 is basically linear between 0.2 Pa and 0.4 Pa. The data in fig. 7 were obtained during a plasma pulse in which gas pressure was linearly ramped; RF input power was 4x60 kW while the filter field current was 750 A; each dot in the plot represents a measurement from a single CRDS signal as in fig. 5.

During the experimental campaign it was not possible to systematically study the influence of total RF power on negative ion density. It was also possible to selectively activate the four RF generators (RF1, RF2, RF3, RF4), each of them powering a row of RF drivers. Fig. 4 indicates with large orange circles the projections of RF drivers with respect to the PG, and gives the correspondence between RF generators and rows of drivers. In absolute terms, RF drivers have an inner diameter of 27.5 cm and the distances between their axes is 40 cm horizontally and about 44 cm vertically; their exit is about 24 cm far from the PG. Due to technical issues, for most of the time it was not possible to deliver RF power from RF1 to its two associated drivers. In a shot with maximum 3x80 kW input power (RF3+RF4+RF2), 0.19 Pa source pressure and 1.3 kA magnetic filter field current, $(8.8\pm0.7)\cdot10^{15}$ m$^{-3}$ D$^-$ were measured at the PG by CRDS. By switching RF3 off, D$^-$ density dropped to $(5.5\pm0.6)\cdot10^{15}$ m$^{-3}$ (-38 %). At last, by also switching RF4 off, i.e. leaving only RF2 active, D$^-$ density was reduced to $(3.7\pm0.6)\cdot10^{15}$ m$^{-3}$ (-58 % with respect to the original case). This test was also performed at 3x100 kW input power, finding similar results. These results suggest that the plasma produced in the drivers can expand over a large area of the PG. It was also noticed that, independently from which RF generators are activated, in a range of 0.2 Pa-0.4 Pa, a minimum total RF power between 120 kW and 180 kW is necessary to have a sufficient density of D$^-$ at the PG to be detectable by the CRDS diagnostic.

During the experimental campaign it was possible to check whether the magnetic filter field action, apart from reducing the amount of co-extracted electrons, has an impact on the production of negative ions in proximity of the PG. By setting the filter field current to 750 A, 1300 A and 2000 A D$^-$ densities of $(3.1\pm0.7)\cdot10^{15}$ m$^{-3}$, $(6.1\pm0.7)\cdot10^{15}$ m$^{-3}$ and $(5.4\pm0.7)\cdot10^{15}$ m$^{-3}$ were measured, respectively. RF power was 3x80 kW, source pressure was 0.19 Pa. A basic interpretation of the results is that the magnetic filter is beneficial in suppressing the electron stripping reactions, however at excessive levels it may slightly depress the volume reactions that lead to the production of negative ions.

## V. CONCLUSIONS

A CRDS diagnostic on the negative ion source SPIDER was designed, installed and is currently routinely available for plasma operation. Despite some concerns due to the drift and the oscillations of the measured ring down time, which are going to be investigated in future, the CRDS in SPIDER is already able to detect negative ion density values down to $5\cdot10^{15}$ m$^{-3}$ in Cs-free operation, for plasma pulses of hundreds of seconds. The measured values of negative ion density are, as expected, one order of magnitude lower than those detected in other sources with Cs evaporation.[5,8,15] The CRDS has already started contributing to the characterization of SPIDER source, in order to maximize the current density that can be extracted from it. The first characterization of the D$^-$ production has clearly shown the dependency of D$^-$ production on, in order of importance, gas pressure, RF power and magnetic filter field current. In the future, three more LoSs will be added to study the uniformity of negative ion production along the main direction of plasma drifts (i.e. in vertical direction). Additional efforts will be made to improve the mechanical stability of the optical setup and to identify and remove the causes of drifts and oscillations in τ. This will allow to improve the diagnostic sensitivity and reliability.

## VI. ACKNOWLEDGMENTS

The work leading to this publication has been funded partially by Fusion for Energy. This publication reflects the views only of the authors, and F4E cannot be held responsible for any use which may be made of the information contained therein. The views and opinions expressed herein do not necessarily reflect those of the ITER Organization.

This paper is dedicated to the memory of Vannino Cervaro, one of the co-authors, who sadly passed away in February 2021.

## VII. DATA AVAILABILITY

The data that support the findings of this study are available from the corresponding author upon reasonable request.